\documentclass[english,aps,prstper,reprint,showpacs,titlepage,longbibliography]{revtex4-2}   

\usepackage[T1]{fontenc}	% should generally be included for better accented-word behavior
\usepackage[latin9]{inputenc}	% should generally be included for better accent behavior
\usepackage{geometry}		% for controlling page margins
\usepackage{graphicx}
\usepackage[above,below]{placeins}	% allows use of \FloatBarrier command to force section barriers
\usepackage{times}
\usepackage{comment}

\usepackage{hyperref}  
\hypersetup{colorlinks=true,urlcolor=blue,citecolor=blue,linkcolor=blue}   
\usepackage{enumitem}          % package for handling list formatting
\setlist{nosep}                 % Tightest spacing for lists. `noitemsep` is more relaxed

\begin{document}
\begin{titlepage}
  \title{Analyzing Physics Majors' Specialization Low Interest Using Social Cognitive Career Theory}
  \author{Dina Zohrabi Alaee (she/her)}
  \affiliation{Physics Department, Rochester Institute of Technology, 84 Lomb Memorial Dr, Rochester, NY,
    14623-5604}  
    \author{Keegan Shea Tonry (they/them)}
  \affiliation{Physics Department, Rochester Institute of Technology, 84 Lomb Memorial Dr, Rochester, NY,
    14623-5604} 
  \author{Benjamin M. Zwickl (he/him)}
    \affiliation{Physics Department, Rochester Institute of Technology, 84 Lomb Memorial Dr, Rochester, NY,
    14623-5604} 
  \begin{abstract}
As students pursue a bachelor's degree in physics, they may ponder over which area to specialize in, such as theory, computation, or experiment. Often students develop preferences and dislikes, but it's unclear when this preference solidifies during their undergraduate experiences. To get a better understanding, we interviewed eighteen physics majors who were at different stages of their degree regarding their interest in theory, computation, and experimental methods. Out of the eighteen students, we chose to analyze only nine students who rated computation and theory the lowest. Our analysis did not include interest in experiment because the ratings were less negative. We used Social Cognitive Career Theory (SCCT) and Lucidchart to analyze students' responses and create individual graphical representations of the influences for each student. Through this, we uncovered how various factors such as learning experiences, self-efficacy, and outcome expectations influenced their low interest in a particular method. We found that lack of knowledge and experience is often the main reason why self-efficacy was lower. Students' lack of interest is also influenced by negative outcome expectations (e.g, math-intensive and a bad work-life balance) more than other SCCT factors. Our findings could help physics departments and educators identify positive and negative factors that could lead to a more motivating and inclusive physics curriculum. %Using these results, ultimately, we aim to develop an interest survey that can gather broader insights from physics majors across the country using the results we've found. 
    \clearpage
  \end{abstract}
    \maketitle
\end{titlepage}
\section{Introduction}
In order to provide students with the best possible education, physics departments should prioritize creating a learning environment that is engaging and challenging and also emphasizes the importance of different method specializations such as theory, computation, and experimental skills. By doing so, students can develop a deeper interest in these areas and gain a better understanding of their future goals and their career options. Students' actual experiences as physics majors may fall short of this goal. For example, some students may not have access to much computational coursework, and most students will get far more practice with theoretical problem solving than with hand-on experimental work.

According to statistics from the American Institute of Physics~\cite{american_institute_of_physics_employment_2020}, 50~\% of physics bachelor's degree recipients who seek jobs in the private sector are working in computing and engineering-related fields, but there isn't much research on what factors influence their career choices. Substantial prior work has focused on how students' physics identity~\cite{hazari2010connecting} and out-of-class learning experiences in high school affect their decision to study physics in college~\cite{dabney2012out, lock2019impact}. Other work has studied interest in physics and long-term retention~\cite{Franklin_2023}. %Our own group 
%-----Beside, Bennett and Cardona has studied career interest formation, particularly around methods of physics \cite{Cardona_2021} and subfields (e.g., biophysics, astronomy) \cite{Bennett_2022}. 

%Physics departments have always been known for their dedication to providing students with the best and highest quality education, with the aim of preparing them for successful careers in physics. They attempt to create learning environments that are engaging and challenging, and that foster curiosity and critical thinking. Through coursework, hands-on research opportunities, and collaboration with professors and peers, students are able to gain the knowledge, skills, and confidence they need to develop an interest in a specific method specialization of physics such as theory, computation, and experimental. %All physics graduates need to be well-prepared for successful careers, whether they choose to pursue opportunities in academia, industry, or government. 
The focus of this study is to gain insight into the factors that affect the career transition of physics graduates, with a specific focus on what may lead to a lack of interest in the core methods of doing physics. The main research question addressed in this study is: \textit{What factors and experiences contribute to physics majors developing a low interest in two particular method specializations (theory and computation) in physics?} By exploring this topic, we hope to better understand the decision-making processes of physics students and identify strategies that support rather than hinder their future career paths.
% It is about understanding the kinds of experiences and factors that influence low interest (as opposed to high interest).
\section{Background}
\begin{figure}[htb]
        \includegraphics[trim=75 149 502 130,clip,width=80mm]{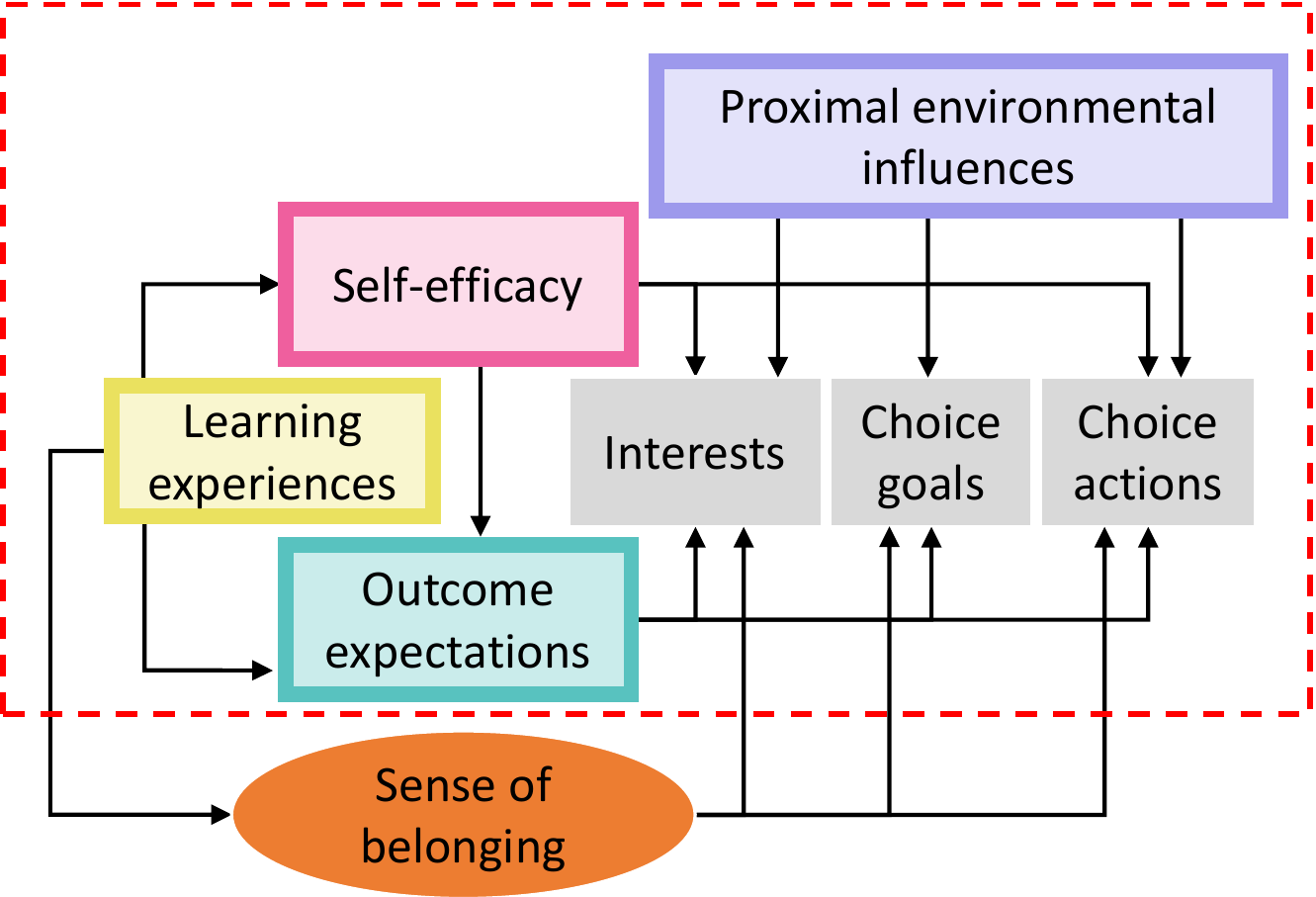}
\centering
%left lower right upper
        \caption{Map of interrelated constructs within Social Cognitive Career Theory and an additional sense of belonging construct.} 
        \label{Fig_1_SCCT.pdf}
\end{figure}

The theoretical framework for this study is based on Social Cognitive Career Theory (SCCT), which reveals different aspects of career and interest development~\cite{lent_longitudinal_2008,lent_toward_1994}. Figure \ref{Fig_1_SCCT.pdf} shows the SCCT map which consists of self-efficacy, outcome expectations, proximal environment influences, interest, and learning experiences. These are the important blocks of the SCCT map that change the most between students. Additionally, the sense of belonging construct is another influential factor in career decisions~\cite{johri_situated_2011, estrada_toward_2011} that we added to the SCCT map.

When considering the development of interest in certain method specializations in physics, it is important to explore what factors and experiences may contribute to this process. SCCT posits that interests are shaped by self-efficacy and outcome expectations that are influenced by learning experiences. Self-efficacy is a student's individual belief in their own capabilities to perform and understand a certain topic, and it is assumed that students gravitate towards fields and topics for which they have stronger feelings of self-efficacy. Having poor performance in previous courses is one example of low self-efficacy among students. On the other hand, outcome expectations are what a student expects work or life will be like as a consequence of working in a particular field or with a specific method specialization. SCCT assumes that people will engage in activities that they expect to be rewarding. One example of a negative outcome expectation is that a particular career path has lots of pressure and a bad work-life balance. Proximal environment influences refer to the contextual variables that influence students' interest. Factors such as family and peer support can have a significant impact on a student's interest. 

Finally, following along the map, choice goals and choice actions refer to an intent or action to take part in activities related to the method of interest and achieve a certain level of success. In our study about low interests, a choice goal or choice action could be avoiding a particular method. Through the exploration of this topic, the study hopes to gain insight into the low-interest formation experienced by physics undergraduate students.

%When considering the development of interest in certain method specializations in physics, it's important to explore what factors and experiences may contribute to this process. According to SCCT, some possible influences may include learning experiences, self-efficacy, and outcome expectations, though it is unclear for physics majors what specifically constitutes a negative outcome expectation or what factors may lead to low self-efficacy. Hence, understanding these factors can provide valuable insights into the formation of low interest in methods of physics. Additionally, external factors such as family and peer pressure, cultural background, and sense of belonging may also play a role in shaping a student's low interest and attitudes toward methods in physics. Figure \ref{Fig_1_SCCT.pdf} shows the SCCT map which consists of self-efficacy, outcome expectations, proximal environment influences, interest, and learning experiences which are the important blocks of the SCCT map that change the most between students. Additionally, the sense of belonging construct is another influential factor in career decisions~\cite{alaee2022impact, johri_situated_2011, estrada_toward_2011} that we added to the SCCT map. %which we recognize it is not completely independent of self-efficacy and outcome expectations~\cite{johri_situated_2011, estrada_toward_2011}.

%SCCT posits that interests are shaped by self-efficacy and outcome expectations that are influenced by learning experiences. Self-efficacy is a student's individual belief in their own capabilities to perform and understand a certain topic, and it is assumed that students gravitate towards fields and topics for which they have stronger feelings of self-efficacy. Having poor performance in previous courses is one example of low self-efficacy among students. On the other hand, outcome expectations are what a student expects work or life will be like as a consequence of working in a particular field or with a specific method specialization. SCCT assumes that people will engage in activities that they expect to be rewarding. One example of a negative outcome expectation is that a particular career path has lots of pressure and a bad work-life balance. Proximal environment influences refer to the contextual variables that influence students' interest. Factors such as family and peer support can have a significant impact on a student's interest. Finally, following along the map, choice goals and choice actions refer to an intent or action to take part in activities related to the method of interest and achieve a certain level of success. In our study about low interests, a choice goal or choice action could be avoiding a particular method. Through the exploration of this topic, the study hopes to gain insight into the low-interest formation experienced by physics undergraduate students.
\section{Methodology}
Our study involved semi-structured interviews with undergraduate physics students in the Summer 2022 and Fall 2022. We recruited participants for our study through various channels, including a student organized Discord server for physics majors, an email through the physics department, and visiting a course for first year physics majors. All participants were from the same large private research university in the United States, which has about 200 undergraduate physics majors. In total, we interviewed eighteen physics students who were at different stages of their studies. Out of the 18 students, we chose to analyze the five lowest ratings for computation and the 5 lowest ratings for theory, which included nine different students. Since the overall rating toward the experimental method was high, we did not include it in our analysis. %Demographic information for these nine participants can be found in table~\ref{tab:demographics-participants}. 
Of the $N=9$ participants, $N=5$ described their gender as ``man,'' $N=4$ ``non-binary,'' and $N=1$ as ``woman''. Regarding race and/or ethnicity $N=7$ identified as ``White,'' $N=1$ ``Latino,'' and $N=1$ ``Asian''.  Finally, $N=4$ were in their fourth year, $N=3$ in their third year, and $N=2$ in their first year. Participation incentives were offered in the form of a \$15 gift card. 

During these interviews, the students rated their interest in each method on a scale of 0 to 10, with 0 indicating no interest at all and 10 indicating very high interest. Throughout the paper we use a shorthand notation to indicate interest levels, such as (T-3) to mean a theory interest rating of 3, or (C-5) to mean a computation interest rating of 5. We delved deeper by asking follow-up questions tied to SCCT factors to uncover the various influences on their level of interest. Each interview was recorded, transcribed using Otter.ai, and then errors were corrected prior to analysis. To investigate the formation of students' interests, we were inspired by the phenomenography method~\cite{marton_phenomenography_1981, marton_phenomenography_1994} to understand the variation in experiences that led to a low interest in theory or computation. The limited sample size of our study limits our ability to reach `saturation' in data collection and data analysis.%To investigate the formation of the student's interests, we used the phenomenography method to obtain a complete understanding of how various ways students described their low interests in theory or computation methods. We also used a concept map tool called Lucidchart~\cite{noauthor_Lucidchart_2023} to create individual flowcharts for each student, allowing us to organize and connect the various factors they explained. 
\begin{figure}[h]
\centering
\includegraphics[trim=116 90 375 70,clip,width=84mm]{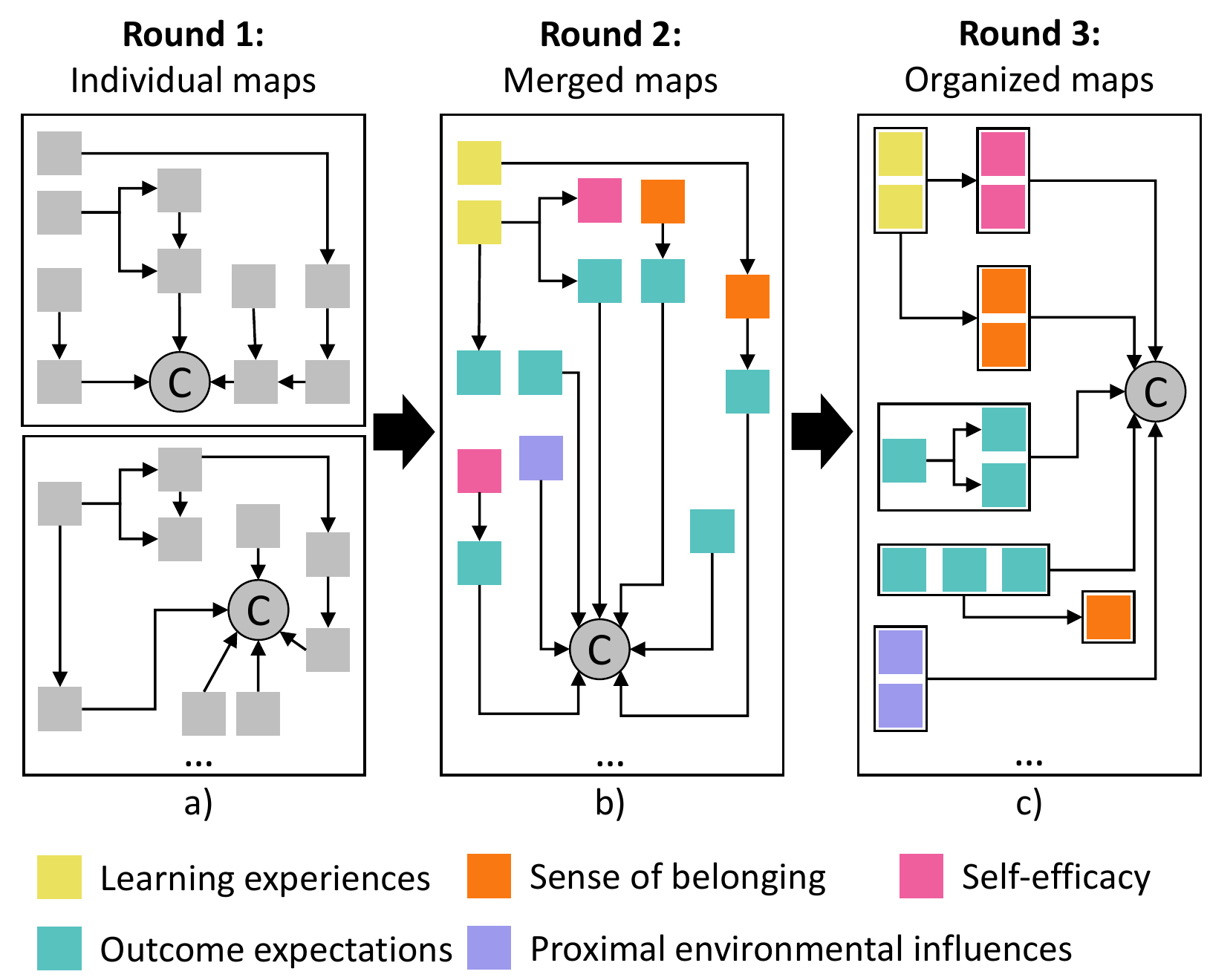}
\centering
%left lower right upper
\caption{Data analysis steps. (C) means a computation.}
\label{Fig: Fig_2_Method.pdf}
\end{figure}
First, we identified each student's meaningful statements and created an individual diagram of each student's influences by using Lucidchart~\cite{noauthor_Lucidchart_2023} (see Fig.~\ref{Fig: Fig_2_Method.pdf}-a)). After creating 10 separate maps, we coded each student's statements into SCCT categories such as learning experiences, self-efficacy, and outcome expectations, as well as sense of belonging. Then, we merged individual maps together without students' quotes and created separate comprehensive maps for theory and computation method (see Fig.~\ref{Fig: Fig_2_Method.pdf}-b)). This allowed us to see the variation in students' experiences and highlighted similar and contrasting factors that impacted their interest formation. Finally for each method, emergent subcategories were identified within the SCCT constructs and sense of belonging factor (see Fig.~\ref{Fig: Fig_2_Method.pdf}-c))).
\vspace{-3mm}
% % % % make a demographic table if we have space if not put it in the context as a text

\section{Results} %\vspace{-3mm}
\subsection{Interest Levels}
We characterized interest based on students' numerical ratings and their qualitative comments. Based on numerical ratings, students had a more negative perception of theory (T) (ratings: 1,2,2,2,4; $\textrm{Av.}=2.2$) compared to the computation (C) (ratings: 3,3,5,5,5; $\textrm{Av.}=4.2$). For the other 13 participants, the average rating was 7.4 for theory (ranging from 6-10) and 7.0 for computation (ranging from 5-10). The average for experimental work was 8 (ranging from 5-10). 

Students' interest was also indicated by their language when describing a particular method. Some students expressed strong negative emotions and language when discussing their lack of interest in theory. For example, Skyler who rated theory a 2 (T-2), stated, ``I hated the materials. It wasn't necessarily a negative class experience, I just did not enjoy the material in the slightest... Just cringe... The professors are all crazy and I'm blanking on every class I've ever taken. Modern was garbage.'' Skyler continued that they had a supportive professor that tried to help them, but they were not paying attention. They said, ``I voluntarily excluded myself from theoretical areas and feel dread to get involved in this type of work.'' Armond (T-1) also used strongly negative language, describing theory as ``miserable,'' ``class was really just gross, ''data is just basically irrelevant,'' and ``Astro stuff was horrible''. Other students who rated theory low had less emotional responses, with Sean (T-2) and Wren (T-2) using words such as ``a little disappointed,'' ``not fun,'' ``not entirely bad,'' and ``not horrible.''

On the other hand, students who rated computation as their low-interest method used a mix of positive and negative language. %One of the reasons why a student may not choose a computation path is because they don't like coding in general.
For instance, Bert (C-3) expressed that seeing programs run efficiently is really cool, but he doesn't enjoy coding. He also admitted that he will need to use some computation physics throughout his professional life, but he prefers to limit it as much as possible. Jamie (C-5) thought ``There's less funkiness in computing, and slightly more funkiness in experiments'', while Moss (C-5) expressed having a love-hate relationship with it since ``it has been frustrating working with computational physics,...but useful.'' %Some students acknowledged the importance of computational work in their chosen profession, making it difficult to completely avoid it. For example, Moss (C-5) said, ``I'm going to be an astronomer or an astrophysicist. It's literally impossible to avoid computation. I can't go out into space and observe things. Essentially, I'm observing all my work through a screen, using programs, software, and coding.'' Many factors that influence students' interest levels are discussed in the following sections.
%Some students used negative words to express their negative thoughts about the method.
\subsection{Learning Experiences}
Within our data, students who were not interested in theory often had negative experiences with theory-related courses, both prior to and during college. Wren's (T-2) low interest in theory was linked to their perception of high school geometry. They said ``Geometry class in high school was a lot of proofs and I just did all the geometry proofs and I was like, I don't like doing proofs.'' 
Additionally, one student shared his negative research experience as a contributing factor in his low interest. Armond (T-1) had a summer research experience in Astrophysics, which influenced his negative perception of theory. He said, ``I didn't really enjoy it. I really liked Astro before that. I was like `Astro is so cool, I'm going into Astro.' But after I did that project, I completely got turned off by Astro. I was like, `I don't want to do this ever again.' The data is just basically irrelevant to me, it has no significance... It's just numbers.'' For Armond, a dislike of theory is coupled to a dislike of astrophysics. %Armond talks negatively about theory, but it is ``data'' that feels irrelevant. 
We will come back to the influence of irrelevance in the section on outcome expectations. %It's important to note that negative perceptions of theory could be linked to students' negative experiences with computation as well, as seen in Armond's low rating for computation (C-3). %Skyler (T-2) also mentioned that his professors for these courses made it worse for him. 

However, we found that students who rated computation lower often had negative experiences with coding during their research experiences. A few students also shared stories about their experiences with computational courses at various levels of education, from pre-college to BS/MS programs. %For example, Dave (C-5) said, ``The research idea is cool but the way [my mentor] went about it in the methods, I think was the biggest disinterest to me. All the research is done using MATLAB there is no apparatus, and there is no model. We're trying to prove we're trying to just discover physics using computational methods.'' 
%However, we found that some students who rated computation lower mentioned that their research experiences involving coding were their significant learning sources. These experiences were in addition to graduate and undergraduate computational courses taken during college and pre-college computer classes. 
It is important to note that not all of these learning experiences were necessarily negative or unpleasant. As we delve deeper into the results, we'll examine what specifically made these experiences unpleasant in subsequent subsections.
%Armond (C-3) took a graduate course as a part of the BS/MS program.
\subsection{Self-efficacy} 
When students were asked about their confidence in a method specialization that didn't interest them, they often reported feeling capable to some extent. For example, Armond (T-1 \& C-3) and Sean (T-2) felt confident in their understanding of physics. Negative perceptions of theory work, as seen in Armond's case, could be linked to his negative experiences with computation but he thought ``If I given the right tools, I could probably do it. I might be bored.'' Kennedy (T-4) and Jamie (C-5), both rising 1st-year students, shared a similar sentiment that they were to some extent confident but also acknowledged that they still had more to learn. For example, Kennedy said, ``I think we'll get more confident over time as I understand more physics laws and stuff like that and be able to apply them to new solutions.'' Dave (C-5) who became a physics major because he wanted a better understanding of the world, said, ``If I actually tried, it could probably be pretty achievable. I would just learn how to code and then learn how to make predictive models.'' However, for Bert, the mathematical process and coding were a particular challenge, and his confidence level varied depending on the complexity of the process. 
%he understands general processes and easily gets through them, but a lot of the time, especially with a higher-level math process, he felt less confident. 
He said, ``I had no clue how to go about coding, how to learn about it, how to do it, how it could improve, how I code, or what I could see that was wrong... Very minimally. I don't understand them. In every other STEM category,...I can at least see how I did something wrong, whereas, for coding, there are no indicators of how your code is wrong. There is but like, very minimally. I don't understand them. My impression of coding was I can't do this without a teacher or I don't have enough motivation to do this without a teacher. Because I tried learning coding on my own at one point, and it just didn't work. Either I just chose the wrong classes or just didn't understand.'' 

A student's self-efficacy in a method specialization can be influenced by pedagogy. For example, Moss (C-5) mentioned feeling more confident in one-on-one teaching situations, however, even in those settings, professors can negatively influence confidence if the student perceives the professor is down-talking or lecturing them. %Armond (T-1 \& C-3) and Sean (T-2), on the other hand, felt confident in their understanding of physics. Negative perceptions of theory work, as seen in Armond's case, could be linked to his negative experiences with computation but he thought ``If I think given the right tools, I could probably do it. I might be bored.''

\subsection{Outcome expectations: Nature of the work} 
Our research has identified three categories of outcome expectations that impact students' interest development: disciplinary knowledge, practices, and professional and personal life. 
\label{Sec: Nature of the work}\\
\textbf{Disciplinary ideas} Disciplinary idea refers to the in-depth content knowledge of a certain type of method specialization in physics. For example, Armond (T-1) who prefers ``hands-on work rather than use his brain'', had difficulty grasping the micro-scale of the theory. Some students, such as Moss (C-5), had less interest in pursuing theory or computation because they were interested in the immediate translation to real-world applications rather than the acquisition of fundamental skills and knowledge. 
%Another negative outcome expectation that students may have of theory and computation paths is that the work won't be useful in real-life experiences and may not directly help people.  
Sometimes in computation and theory, students may struggle with visualizing their work and not feel a physical connection to it. Sean (T-2) described it as ``It can feel like just stringing ideas together'' without anything concrete to hold on to. However, there are benefits to using computation in place of physical experimentation. For instance, it can save time and money. Armond (C-3) argued that by allowing the ``computer to do the heavy lifting'', you just have to give it instructions. This can be challenging, but it can also be more cost-effective than conducting experiments over time.\\ %Plus, you can create and test materials on a computer, eliminating the need for an experimental setting. 
%Armond said The computer's doing most of the heavy lifting for you. so you just have to tell it what to do, which is hard... I mean, you can make the material on a computer and test it on a computer, rather than having to do it in an experimental setting, and that's going to save you tons of money if you like, do it like over time. 
\textbf{Practices} 
Outcome expectations in the practices subcategory refer to the day-to-day performance of a certain skill or task associated with each type of method specialization in physics. It is common for students to think that both computation and theory are math-intensive methods. Wren (T-2) feels that ``theory involves a lot of proof'' while Sean (T-2) finds it ``unsatisfying to sit and do derivations.'' Additionally, some students believe that these types of work won't involve collaboration, as Dave (C-5) prefers ``interacting with people and seeing the progression of hard work.'' There is also a negative expectation that theory and computation work won't be hands-on and will involve lots of sitting in front of the computer and lots of coding and formalizing ideas. For example, Wren (T-2) likes having their own thoughts but dislikes having to mathematically formalize them. Additionally, some students find working in theory, to be ``time-consuming,'' with ``lots of Zoom calls,'' ``long hours [of] reading,'' ``lots of literature searching and lots of generative writing,'' ``understanding textbooks,'' ``using a lot of chalkboards,'' and ``justifying each [idea].'' For instance, Kennedy avoids theory because she ``likes to figure things out quickly.''

\subsection{Outcome expectations: Professional and personal life}

%Kennedy (T-4): “I want to have like a healthy balance between work and social life and my other hobbies. But I feel as if I was constantly [have] stuff in my head. I'd bury my head in it.”
In contrast to the outcome expectation subcategory for practices, which was based on perceptions of the nature of the work and day-to-day activities, this section refers to the outcome expectations in terms of lifestyle. As students think about their future careers, they are considering their lifestyle expectations and how they can find a balance between their professional work and personal life. They want to prioritize their mental well-being and create a healthy work-life balance. Kennedy (T-4), for example, hopes to have time for ``social life and my other hobbies'' while avoiding the stress of constantly thinking about work. %Moss (C-5) was interested in the flexibility that comes with working remotely or traveling for conferences. 
While Moss (C-5) had a love-hate relationship with computation (i.e.,frustrating but useful), they still see computation work as a way to align their work with their desired lifestyle, which is working remotely or traveling for conferences. They said, ``when you're a researcher, that's essentially what I want to be, and when [you are] doing research hopefully it would require me to travel sometimes.'' %Ultimately, students are looking for a career that will allow them to pursue their passions and enjoy a fulfilling life outside of work. 
According to students, working in theory means having an office in a department and working as a professor with students. 

\subsection{Proximal environment influences}

Some students' interests in computation may be influenced positively or negatively by environmental factors such as family or friends in the field. Moss (C-5), for example, was initially pressured by their father to pursue computer science, but the pressure actually deterred them from it as a teenager. However, as students' interests can evolve over time, encouragement and support from loved ones can have a significant impact in reshaping their future career paths. In addition to being a role model for their little brother at home, Moss explained how they work in a nursing home and have opportunities to explain their research to ``old folks'' who are  ``completely impressed and proud''. Besides family and friends, Armond (C-3) and Dave (C-5) found it challenging to interact with their learning community during the limitations created by the COVID-19 pandemic, which lowered their interest. 
%By recognizing this and providing the necessary support, academic departments can help their students achieve their goals and reach their full potential. It's heartening to see examples like Moss, who is impressing and making their family proud by working at a nursing home, while also being a role model for their little brother. Let's continue to focus on these important areas to provide new insights and improve student outcomes.

\subsection{Sense of belonging} 

Sense of belonging is an important element that is not part of the SCCT framework and focuses on the social connections of students within their community. Creating a sense of connectedness and being recognized by others through the different learning experiences can enhance interest in a particular method. Moss (C-5) shared their experience of feeling out of place in the computation community, where they rarely felt encouraged or congratulated. Moss also noticed a significant gender imbalance in computation classes, which made them feel like they didn't belong in those spaces. However, they found more support in their research setting, where their peers and mentor provided them with more moral support, leading to a more positive experience with computational physics. Moss said once they did work, their peers are ``like, Oh, my God, congrats, good job! They all send a little reaction, like a cute congratulations emoji and that kind of stuff is important.'' On the other hand, if you don't have these interactions it can lead to a lack of sense of belonging as seen in Sean's case where he felt excluded in theory classes because his peers didn't take him seriously. In addition, Sean (T-2) also stated that he has ``to prove myself to the professors by the way that I look.'' 
\section{Conclusion}
In our sample, we noticed that students had a generally more negative view of theory compared to computation based on the ratings and emotional language. Some students acknowledged the importance of computational work in their chosen profession, making it difficult to completely avoid it. For example, Moss (C-5) said, ``I'm going to be an astronomer or an astrophysicist. It's literally impossible to avoid computation. I can't go out into space and observe things. Essentially, I'm observing all my work through a screen, using programs, software, and coding.''
%Additionally, computation plays a vital role in most careers today, making it hard to avoid. 
On the other hand, students perceived theory as irrelevant and not useful in their daily lives. Students consistently described the theory as difficult. 

When it comes to interests, there are a lot of factors that come into play. Some of these factors are positive, while others are negative, and some might outweigh others. For the 1$^\textrm{st}$year students, lack of knowledge and experience is often the main reason why their self-efficacy is fairly low. Negative outcome expectations tended to have a bigger impact on students' lack of interest than other SCCT factors like self-efficacy. Students with positive self-efficacy might still show low interest if the outcome expectation is negative. 

We identified three emergent categories of outcome expectations that impact students' interest development: disciplinary knowledge, practices, and professional and personal life. By understanding how each of these outcome expectations categories affects students' interests, we hope to develop strategies that can guide students toward accurately assessing whether their goals and outcome expectations align. 

As a limitation, we do not have evidence that our themes achieved saturation. We intend to do a full analysis of all 18 interviews and all three method specializations (including experiment). Our long-term goal is to design departmental assessment tools that can identify which factors contribute most positively and negatively to students' interest in a particular method of doing physics. We hope this work will lead to evidence-based strategies that can be implemented in physics departments to support students in their academic and professional journeys. 

%\section{acknowledgments}
%This work was supported by the National Science Foundation under Grant No. 1846321.

% \bibliographystyle{apsrev} % supercedes the longbibliography option, so leave commented out if you want to display article titles

%\citeanon[option]{bibentry}
\bibliography{bibfile} % don't include the .bib suffix
\end{document}